%% file: main.tex
\documentclass[conference]{IEEEtran}
\IEEEoverridecommandlockouts
\usepackage{cite}
\usepackage{amsmath,amssymb,amsfonts}
\usepackage{algorithmic}
\usepackage{graphicx}
\usepackage{textcomp}
\usepackage{float}
\usepackage{xcolor}
\def\BibTeX{{\rm B\kern-.05em{\sc i\kern-.025em b}\kern-.08em
    T\kern-.1667em\lower.7ex\hbox{E}\kern-.125emX}}
\begin{document}

\title{BioTrak: A Blockchain-based Platform for Food Chain Logistics Traceability\\

}

\author{\IEEEauthorblockN{1\textsuperscript{st} Alessia Spitalleri}
\IEEEauthorblockA{\textit{PeRCeive Lab} \\
\textit{University of Catania}\\
Catania, Italy \\
alessia.spitalleri@unict.it}
\and
\IEEEauthorblockN{2\textsuperscript{nd} Isaak Kavasidis}
\IEEEauthorblockA{\textit{Istituto Nazionale di Astrofisica} \\
\textit{ INAF Catania}\\
Catania, Italy \\
isaak.kavasidis@inaf.it}
\and
\IEEEauthorblockN{3\textsuperscript{rd} Vincenzo Cartelli}
\IEEEauthorblockA{\textit{ALLinCLOUD Srl} \\
\textit{University of Catania}\\
Catania, Italy \\
vincenzo.cartelli@allincloud.eu}
\and
\IEEEauthorblockN{4\textsuperscript{th} Raffaele Mineo}
\IEEEauthorblockA{\textit{PeRCeive Lab} \\
\textit{University of Catania}\\
Catania, Italy \\
raffaele.mineo@unict.it}
\and
\IEEEauthorblockN{5\textsuperscript{th} Francesco Rundo}
\IEEEauthorblockA{\textit{ADG, R\&D Power and Discretes} \\
\textit{ST Microelectronics, Catania, Italy}\\
francesco.rundo@st.com}
\and
\IEEEauthorblockN{6\textsuperscript{th} Simone Palazzo}
\IEEEauthorblockA{\textit{PeRCeive Lab} \\
\textit{University of Catania}\\
Catania, Italy \\
simone.palazzo@unict.it}
\and
\IEEEauthorblockN{7\textsuperscript{th} Concetto Spampinato}
\IEEEauthorblockA{\textit{PeRCeive Lab} \\
\textit{University of Catania}\\
Catania, Italy \\
concetto.spampinato@unict.it}
\and
\IEEEauthorblockN{8\textsuperscript{th} Daniela Giordano}
\IEEEauthorblockA{\textit{PeRCeive Lab} \\
\textit{University of Catania}\\
Catania, Italy \\
dgiordano@unict.it}

}

\maketitle

\begin{abstract}
\input{abs.tex}
\end{abstract}

\begin{IEEEkeywords}
Food supply chain monitoring, Business Process Management, Blockchain
\end{IEEEkeywords}

\section{Introduction}
\input{intro.tex}

\section{Related work}
\input{rel.tex}

\section{The BioTrak Platform}
\label{sec:3}
\input{app.tex}

\section{Conclusion}
\input{conc.tex}

\section*{Acknowledgment}

\input{ack}

\bibliographystyle{IEEEtran} 
\bibliography{bib} 

\end{document}

%% file: abs.tex
The food supply chain, following its globalization, has become very complex. Such complexities, introduce factors that influence adversely the quality of intermediate and  final products. Strict constraints regarding parameters such as maintenance temperatures and transportation times must be respected in order to ensure top quality  and reduce to a minimum the detrimental effects to public health. This is a multi-factorial endeavor and all of the involved stakeholders must accept and manage the logistics burden to achieve the best possible results.

However, such burden comes together with additional complexities and costs regarding data storage, business process management and company specific standard operating procedures and as such, automated methods must be devised to reduce the impact of such intrusive operations.  

For the above reasons, in this paper we present BioTrak: a platform capable of registering and visualizing the whole chain of transformation and transportation processes including the monitoring of cold chain logistics of food ingredients starting from the raw material producers until the final product arrives to the end-consumer. 

The platform includes Business Process Modelling methods to aid food supply chain stakeholders to optimize their processes and also integrates a blockchain for guaranteeing the integrity, transparency and accountability of the data.

%% file: intro.tex
The food supply chain is a complex network of producers, distributors, retailers, and consumers that is responsible for the production and distribution of food around the world. However, the food supply chain is also plagued with a number of challenges such as food fraud, safety concerns, and inefficient supply chain management. These issues can have serious consequences for public health, the environment, and the economy 
\cite{bhat2021agriculture,kamilaris2019rise}. 

To address these challenges, there is a growing interest in the use of blockchain technology in the food supply chain. Blockchain is a decentralized and secure digital ledger that allows for the transparent and immutable recording of transactions. This technology has the potential to improve the transparency, traceability, and accountability of the food supply chain, thereby enhancing food safety.

Blockchain technology can be used to track and record the movement of food products from farm to table, providing a complete and transparent record of each transaction. This can help to identify the source of any safety or quality issues, and can also help to prevent food fraud by ensuring that products are properly labeled and authenticated.

In cases such shortcomings have no catastrophic consequences, warranties must be offered to the next step in the food supply chain sequentially until reaching the top level (i.e., the final consumer).

In this work we present the BioTrak platform: an application aiming at gathering and organizing implicit and explicit feedback about food transformation and showing relative information to the end-consumer in a user-friendly graphical interface. The application was designed to be as less intrusive as possible to the whole food supply chain's operators in order to not interfere with their activities. 

The platform integrates blockchain technology to monitor food supply chain and ensure their transparency, traceability, and accountability by promoting or offering:
\begin{itemize}
    \item Product origin tracking by assigning a unique identifier and tracked through the supply chain from the farm to the store shelves. Information about the product, such as its origin, production date, and storage conditions, can be recorded and stored on the blockchain.
    \item Quality assurance by recording information about various aspects of the food supply chain, such as temperature and humidity levels during storage and transportation. This information can be used to verify that food products are being stored and transported under the proper conditions to ensure their freshness and safety.
    \item Traceability by quickly tracing the origin of the contaminated product and identify all other products that may have been affected. This can aid for prompt mitigating action and contain the problem and prevent it from spreading.
    \item Supply chain efficiency by providing real-time information about product availability, reducing the need for manual data entry and reducing the risk of errors.
    \item Enhanced transparency by creating a transparent, secure and tamper-proof record of the food supply chain, all stakeholders can have confidence in the information being recorded, and in the food products they are purchasing.
\end{itemize}

Additionally, the platform offers tools for monitoring cold transportation parameters by integrating methods capable of interfacing with NFC temperature sensors. 

The remainder of this paper is as follows: the next section presents a literature analysis of current solutions for food supply chain monitoring, while Section \ref{sec:3} presents the BioTrak platform and its constituent parts. Finally, in the last section conclusions are drawn and future directions are given.

%% file: rel.tex
During the last decade, the invention of the blockchain\cite{nakamoto2008bitcoin} as a concept and its adoption and further research and application to many domains provided the solution to a number of problems regarding data integrity, traceability and accountability. 

Indeed, blockchain-based technologies have been adopted in many  sectors where such data qualities are necessary, such as in the pharmaceuticals industry \cite{mattke2019enterprise}, in healthcare \cite{agbo2019blockchain} and in finance \cite{guo2016blockchain}. 

The importance of the benefits provided by  blockchain technologies to the food supply chain is already described in the literature and the need for more traceable and transparent information gathered throughout the whole chain is  paramount \cite{feng2020applying,creydt2019blockchain,kramer2021blockchain}.

Many works employ blockchains for food supply monitoring. However the majority of these works are intrusive to the everyday operation of production, transportation or food transformation companies (e.g., \cite{mondal2019blockchain,wang2022design}) or do not integrate sensor data \cite{dey2021foodsqrblock,basnayake2019blockchain}. 

To the best of our knowledge, there not exist any works that combine the features offered by BioTrak: a distributed blockchain-powered platform for easy-to-use food supply chain monitoring and an adaptable data format that can integrate a variety of NFC based sensor data.

%% file: app.tex
The BioTrak platform is a set of complementary technologies aiming at gathering, organizing and visualizing information regarding the food supply chain. The central concept of BioTrak is the food intermediate or final product and all data regarding the initial, intermediate and final steps for its creation are grouped together in order to provide a concise and clear view to every interested stakeholder.

The Biotrak platform is developed using the client-server paradigm. However, because the nature of the problem dictates so, a multi-server architecture is adopted. The server application permits its deployment in two modalities, depending on the needs of the stakeholder that hosts it:
\begin{itemize}
    \item Authoritative: Stakeholders that need to write data to the blockchain are needed to register their server instance to the BioTrak network. Such server instances, provide the full spectrum of functionalities (blockchain transaction, business process modeling, sensor integration) and are generally suitable by food production, processing and transportation companies.
    \item Non-authoritative: Stakeholders that need only  to read data from the blockchain do not need to deploy the full set of functionalities. For this reason a lightweight version of the server is also provided that has reduced hardware requirements and is destined to public regulation and safety authorities. However, given that transparency is one of the foremost sought qualities, everyone with a conventional PC and an internet connection can host a non-authoritative server.
\end{itemize}

\subsection{Business Process Modelling}

The supply chain can be thought as a chain of production systems (i.e., seen as a “black box” in which certain input resources from a supplier system are transformed into products for a customer system) starting from raw materials to finished products, sold at retail. Each production process is generally connected to the next by an internal or outsourced transport service which takes care of transferring the output products leaving a company as input resources for the next company in the supply chain.

\begin{figure*}[htbp]
\centerline{\includegraphics[width=16cm]{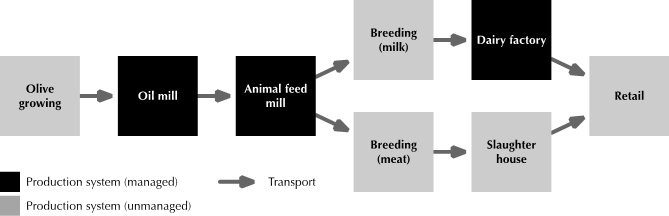}}
\caption{The BioTrak project supply chain. The various  production systems whose processes have been included for internal tracking (managed production systems): the production systems whose processes have been analysed (managed, in black); those that, despite being an integral part of the supply chain, are contributing indirectly  as commercial and non-BioTrak project partners (unmanaged, in grey).}
\label{img:schain}
\end{figure*}

The supply chain shown in Fig \ref{img:schain} has been broken down to a first level of detail for project partner companies in order to identify the technological and informational requirements by analysing the production processes and the IT tools used by each organisation.

\begin{figure}[hbp]
\centerline{\includegraphics[width=9cm]{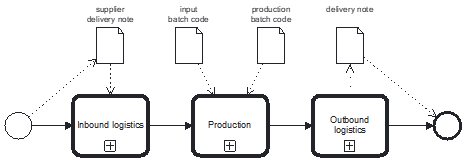}}
\caption{High level representation of the production system.}
\label{img:high}
\end{figure}

All of the partners belong to the production category, so the first level of detail for processes follows the same pattern, with the only deviation in input and output data. This pattern consists of three activities and is shown in Fig \ref{img:high}:

\begin{figure}[htp]
\centerline{\includegraphics[width=9cm]{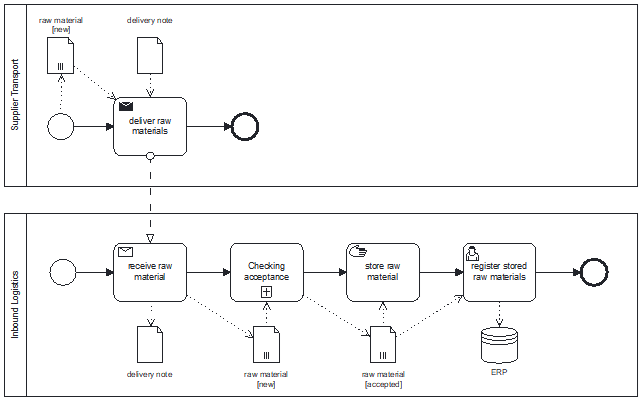}}
\caption{The Inbound logistics process model.}
\label{img:inbound}
\end{figure}

\begin{figure}[h]
\centerline{\includegraphics[width=8cm]{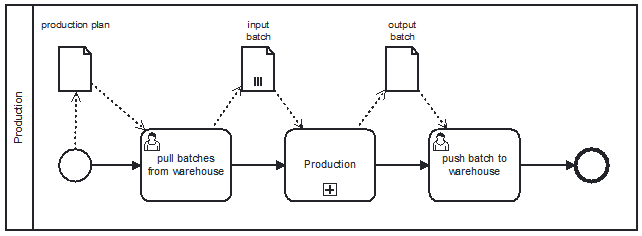}}
\caption{The Production process model.}
\label{img:prod}
\end{figure}

\begin{figure}[h]
\centerline{\includegraphics[width=8cm]{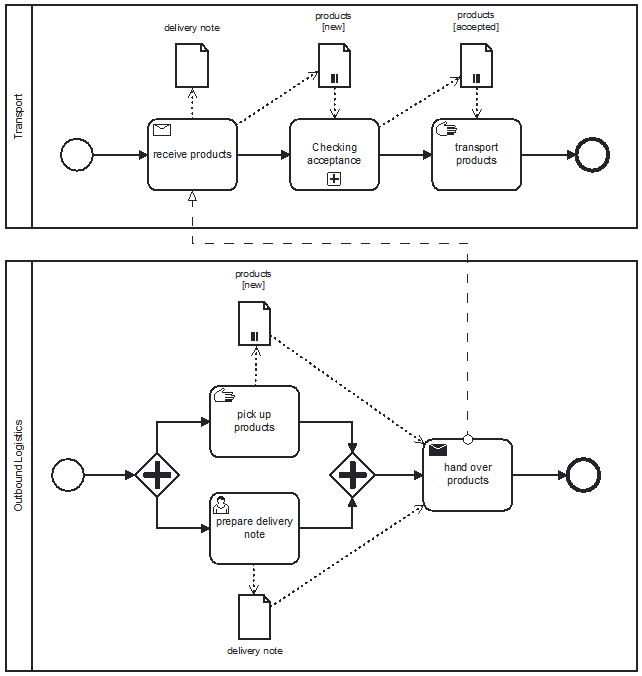}}
\caption{The Outbound logistics process model.}
\label{img:outbound}
\end{figure}

\begin{enumerate}
\item Inbound logistics: raw materials are acquired from a supplier and stored in the corresponding warehouses and moved to the production process when needed.
The supplier delivery note contains a reference to the incoming materials with their respective batch codes (Fig. ~\ref{img:inbound}).

\item Production: raw materials and semi-finished products are transformed – through the production process – in (semi-)finished products for the (internal) consumer.
This is where the (input) resource batch codes are linked to the (output) product batch code (Fig. ~\ref{img:prod}).

\item Outbound logistics: finished and semi-finished products are stored in the corresponding warehouses and moved to the consumer (note that the consumer can eventually be an internal production process in case of semi-finished products) (Fig. ~\ref{img:outbound}).
When the product is supplied to an external consumer, a delivery note containing a reference to the outgoing products with their respective batch codes will accompany the delivery.
\end{enumerate}

Also, as already mentioned, the three processes have been analysed in order to acquire the functional and non-functional requirements for the integration with the supply chain tracking system. This system is implemented through QR code scanning in the BioTrak app and for this reason the delivery notes and batch codes come in the form of QR codes generated automatically by an authoritative BioTrak node.  The second level processes for inbound logistics, production and outbound logistics are illustrated in Fig.~\ref{img:inbound}, Fig.~\ref{img:prod} and Fig.~\ref{img:outbound}, respectively, using the BPMN standard.

\subsection{The blockchain}
\begin{figure}[htbp]
\centerline{\includegraphics[width=8cm]{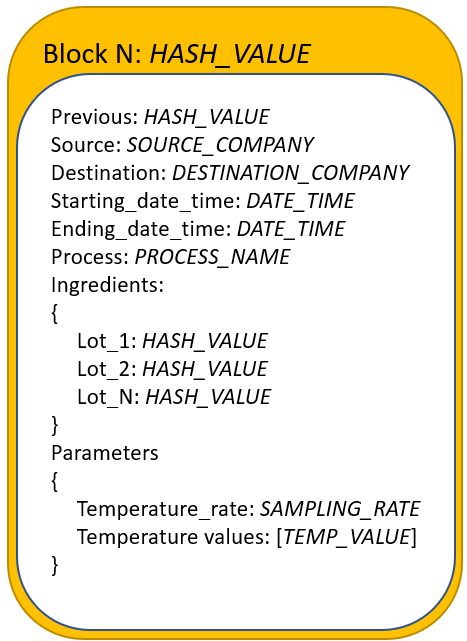}}
\caption{The generic template for each block in the blockchain. The parameters field is flexible and can accommodate every future need in terms of data storage.}
\label{img:bc}
\end{figure}

Guaranteeing the integrity of the information in the BioTrak platform is of paramount importance in order to ensure the best possible quality of the final food products and, for this reason, a customized blockchain is used. At each step in food supply chain where a transportation or transformation process takes place the BioTrak platform registers such events in a blockchain with connections to previous processes that regard the same ingredient lot.  Each block contains a single transaction (i.e., a single transportation or transformation process) that includes all  necessary information of each process and links to other blocks to offer the capability to reconstruct the complete history of processes that led to this block. The information that is registered within each block is shown in Fig.~\ref{img:bc}. 

\begin{figure}[htbp]
\centerline{\includegraphics[width=9cm]{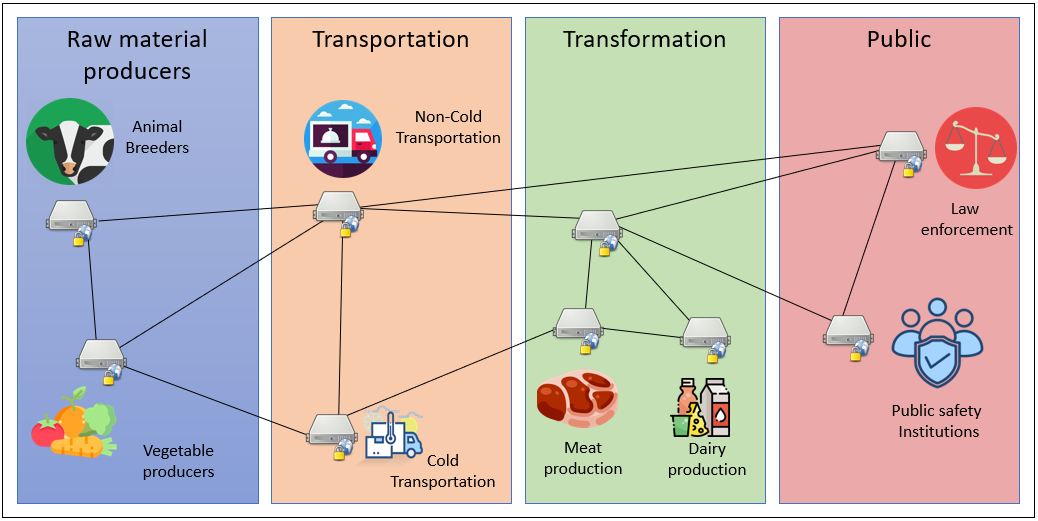}}
\caption{A typical blockchain deployment. All the participants in the food supply chain (raw material producers, transportation and transformation companies) hold an authoritative node (i.e., they can validate and add transactions) of the blockchain operating with a PoA consensus mechanism. Public safety institutions and authorities can hold non-authoritative nodes, ensuring integrity, transparency and accountability of the data. }
\label{img:node}
\end{figure}

\begin{figure}[htbp]
\centerline{\includegraphics[width=9cm]{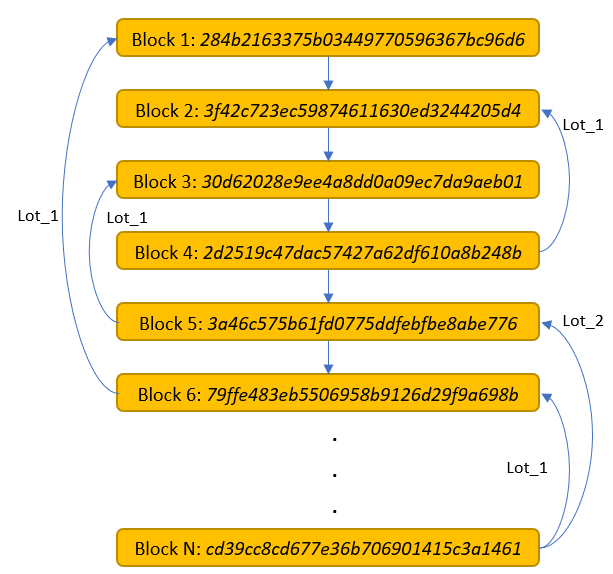}}
\caption{An example of process history  representation in the blockchain. The bottom block (\emph{Block N}) represents the last step in the production. By using the Lot field, connections to the constituent lots and past processes can be retrieved in order to recreate the complete pipeline of transportation or transformation processes. }
\label{img:chain}
\end{figure}

A generic chain of processes regarding a singe food product can be seen in Fig.~\ref{img:chain}

The blockchain operates using a Proof-of-Authority (\emph{PoA}) consensus mechanism. This choice was made because other types of consensus mechanisms available do not fit the context of the BioTrak platform. In fact, the Proof-of-Work consensus mechanism requires unnecessary mining nodes for validating transactions with the subsequent energy expenditures. On the other hand, Proof-of-Stake introduces the ``node importance'' concept that does not fit in a completely distributed peer-to-peer network with equal peers. 

However, given that blockchain validation is entrusted to few centralized nodes, the security of these nodes must be absolutely assured. Also, in order to guarantee unauthorized alterations of the blockchain data, at least 3 transaction validation nodes must be in place and fully operational at any given moment. For this reason, each food production or transportation company that participates in the BioTrak platform needs to host a blockchain node in order to be able to add transactions to the blockchain. 

While adding a transaction in the blockchain regards only the companies involved in the food supply chain, fetching the transaction information is accessible by everyone with the BioTrak application installed in their mobile device. Additionally, public safety institutions are able to deploy \emph{non-authoritative} nodes (i.e., nodes that cannot validate transactions but hold an exact copy of the current state of blockchain) in order such information is needed in food quality investigations. Such feature alleviates public authorities from the time-consuming burden of contacting each individual food transportation or transformation company guaranteeing at the same time the integrity and the transparency of the committed operations.

An example of the blockchain node deployment can be seen in Fig.~\ref{img:node}.

\subsection{The application with sensor data integration}

Except from the server application that was described in the previous subsections, the BioTrak platform offers a mobile app where partners and stakeholders, depending on their roles and permissions, can add new processes happening to food products. 
By using this app, partners can add new processes in food products or update existing ones and all this information is automatically stored in the blockchain. Anonymous users, including final consumers or interested parties can use the app to track all the information regarding the transformation processes of an end product. 

\begin{figure}[htbp]
\centerline{\includegraphics[width=8cm]{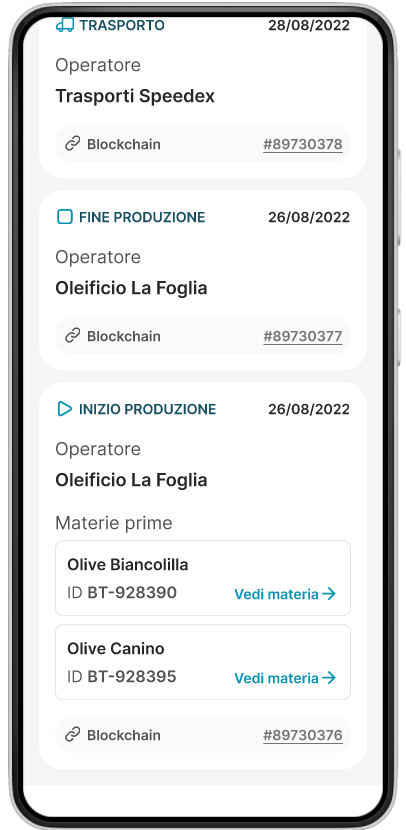}}
\caption{The main GUI of the app that shows the process history of a food product after scanning its QR code. }
\label{img:app1}
\end{figure}

For all the intended uses, the app employs the integrated photo camera of a mobile device to scan the QR code of a product and allows the retrieval of the whole process tree from the blockchain (Fig. \ref{img:app1}).  

Currently, BioTrak supports two roles that grant  rights to write or modify data in the platform:
\begin{itemize}
    \item Producer: A user with this role can register new or update existing food transformation processes. 
    \item Transporter: A user with this role can only register a new transportation event.
\end{itemize}

In addition to the previous two, BioTrak defines also a read-only role (Anonymous) that permits everyone with a mobile device and the app to retrieve quickly information about the food product. This permission is also, by default, granted to the Producer and Transporter roles.

\begin{figure}[htbp]
\centerline{\includegraphics[width=8cm]{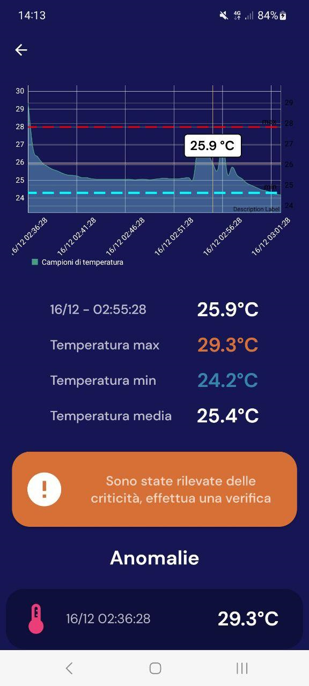}}
\caption{The  mobile  app GUI of the temperature monitoring. }
\label{img:temp}
\end{figure}

The QR codes that can be read by the app are divided in two categories depending on whether the QR code is printed on an NFC-enabled temperature sensor for cold chain monitoring or not. In the former case, the app immediately after reading the QR code, it activates also the NFC feature of the mobile device if present, and downloads the registered data and if asked by the user to visualize it (Fig. \ref{img:temp}). Normally, such feature is used whenever a transportation process is terminated by a recipient. At that point, the transportation process is terminated and an entry together with the downloaded data produced by the temperature sensor is packaged in a block and  appended  to the blockchain. 

%% file: conc.tex
In this work we presented BioTrak, a set of applications and services that offer advanced tracking capabilities in the food supply chain. The application is developed with data transparency and integrity in mind so a distributed blockchain-based mechanism ensures the persistence and access of the data to whoever is interested in it.

The multi-server part guarantees reliability and availability of service and also provides all the necessary tools to stakeholders for monitoring the transportation and transformation processes in the food industry. 

While only the cold supply chain was considered and integrated in the current form of the BioTrak platform, the extensible data format can be exploited to integrate other types of IoT sensors (e.g., humidity \cite{valchanov2022blockchain}).

%% file: ack.tex
This publication has been financially supported by Regione Sicilia through the BioTrak national project (n. 08SR1091000150, CUP G69J18001000007, PO-FESR 2014-2022 Azione 1.1.5)